\documentclass[aps,pre,twocolumn,showpacs]{revtex4}
\usepackage{graphicx}
\usepackage[dvips,
            colorlinks=true, 
            citecolor=blue,
            urlcolor=blue,
            linkcolor=blue,
            ]{hyperref}
\usepackage{amssymb}
\usepackage{bbm}
\newcommand{\abs}[1]{\left\vert#1\right\vert}

\begin{document}

\title{Finite-size effects on open chaotic advection}

\author{Rafael D. Vilela, Alessandro P. S. de Moura, and Celso Grebogi}
\affiliation{Instituto de F\'{\i}sica, Universidade de S\~ao
Paulo, Caixa Postal 66318, 05315-970, S\~ao Paulo, SP, Brazil}


\begin{abstract}
We study the effects of finite-sizeness on small, neutrally 
buoyant, spherical particles advected by 
open chaotic flows. 
We show that, when projected onto configuration space, 
the advected finite-size
particles disperse about the unstable manifold of 
the chaotic saddle that governs the passive advection.
Using a discrete-time system for the dynamics, we 
obtain an expression predicting
the dispersion of the finite-size particles 
in terms of   
their Stokes parameter at the onset of the finite-sizeness induced 
dispersion.
We test our theory in a system derived from a
flow and find remarkable agreement 
between our expression and 
the numerically measured dispersion. 
\end{abstract}
\maketitle


\section{Introduction}

A proper understanding of chaotic advection \cite{aref1}
in incompressible, 
open flows
is substantially relevant, as applications
range from laboratory experiments \cite{horner1} to 
environmental flows of oceanographic \cite{abraham1} 
and atmospherical importance \cite{koh1}. 
Usually, open flows displaying chaotic advection 
are asymptotically regular, 
but the dynamics of the advected particles
possesses a chaotic saddle \cite{kantz1}
which governs the transient
behavior of the orbits. This saddle is an invariant set 
consisting of infinitely many unstable orbits 
organized in a fractal structure. Its stable and 
unstable manifolds also display fractality and can 
be observed directly in physical space. 
In particular, the unstable 
manifold is traced out by an ensemble of fluid particles 
initially placed in the region of the saddle.

The fundamental aspects
of chaotic advection in open flows are now 
relatively well known
(see \cite{karolyi1} and references therein), but almost exclusively  
in the case where the particles are considered to move as 
fluid particles (passive tracers), 
without inertia. 
This situation is termed passive advection.
In many important flows, however, 
the finite-sizeness
of the particles has to be considered \cite{balkovsky1}.
The resulting dynamics is strikingly different
from passive advection. 
In particular, the phase space for finite-size particles has 
twice the number of
dimensions of the phase space of passive advection.
The latter is simply the configuration space (physical space), 
and corresponds to an invariant subspace of the former.
The higher dimensionality  
results from the 
new degrees of freedom corresponding to the
components of the velocity
of the finite-size particles.
This is a consequence of the fact that the finite-size particles 
are not constrained to have the velocity of the advected fluid.  
By contrast, 
the velocities of passive tracers
do coincide with the velocity of the flow.

One consequence of this difference is that 
the motion of finite-size particles
can diverge from the motion of the corresponding 
passive tracers in certain regions of the flow,
even when
the particles have the same density as the fluid \cite{babiano1}. 
We stress that this divergence arises solely from the
finite-size character of the particles, which we consider here to be 
neutrally buoyant \cite{benczik1}, i.e., they
have the same density as the fluid.
The aim of this paper is to investigate the 
consequences of this divergence to open chaotic advection.
We argue and show that, when projected onto 
the configuration space,  
the advected finite-size particles 
disperse about 
the chaotic saddle (and its unstable manifold)
corresponding to passive advection.
This is a general phenomenon that had 
not been reported so far. 
We develop a theory to analyze it.
We obtain a quantitative expression for
the  dispersion
of the finite-size particles about
the chaotic saddle and its
unstable manifold in configuration space
as a function of the Stokes parameter.
We test and validate our theory using the 
blinking vortex-source system and 
find a strikingly good agreement with the 
directly measured dispersion. 
A major consequence of this dispersion is that finite-size effects
can destroy the fractal structure of
open chaotic advection in configuration space, 
with severe consequences to active flows \cite{toroczkai1,nishikawa1}.

\section{Continuous description of the dynamics}

In dimensionless form, the equation of motion for a small rigid spherical 
particle advected by an incompressible fluid of same density with 
a given velocity field ${\bf u}({\bf r},t)$ is \cite{maxey1}
\begin{equation}
\frac{d{\bf v}}{dt}=\frac{D{\bf u}}{Dt}-\frac{1}{\mbox{St}}
\left({\bf v}-{\bf u}\right)
-\frac{1}{2}\left(\frac{d{\bf v}}{dt}-\frac{D{\bf u}}{Dt}\right),
\label{maxey}
\end{equation}
where ${\bf v}$ is the velocity of the particle and
$\mbox{St}=\frac{2a^2 U}{9\nu L}$ is its Stokes number.
Here $a$ is the radius of the particle, whereas 
$U$ and $L$ are the characteristic velocity and lenght 
of the flow, respectively. The kinematic viscosity
of the fluid is given by $\nu$.
Physically, the Stokes number is a measure of the 
finite-size effects.
In the limit of vanishing Stokes number, we recover
passive advection, with ${\bf v}={\bf u}$.
Equation (\ref{maxey}) is Newton's law 
with the terms on the right hand side 
corresponding, respectively, to the force exerted by 
the undisturbed flow, Stokes drag and the added-mass effect.
In this approximation, the Fax\'en corrections
and the Basset-Boussinesq history force term are neglected
\cite{maxey1}.

If we write the derivative along the trajectory of 
the fluid element, 
$\frac{D{\bf u}}{Dt}=\frac{\partial 
{\bf u}}{\partial t}+({\bf u}\cdot\nabla){\bf u}$, 
in terms of the derivative along the trajectory of
the particle, 
$\frac{d{\bf u}}{dt}=\frac{\partial 
{\bf u}}{\partial t}+({\bf v}\cdot\nabla){\bf u}$,
we cast Eq.  (\ref{maxey}) into  the form \cite{babiano1}
\begin{equation}
\frac{d{\bf A}}{dt}=-\left(\mathsf{J}_{\bf u}+\frac{2}{3\mbox{ St}}\mathbbm{1}\right)
\cdot{\bf A},
\label{babiano}
\end{equation}
where ${\bf A}={\bf v}-{\bf u}$ and $\mathsf{J}_{\bf u}$ is the Jacobian 
of ${\bf u}$.
For convenience, we treat a two-dimensional fluid flow, 
so the dynamics of finite-size particles 
occurs in a $4$-dimensional  phase space. The configuration 
space, corresponding to ${\bf A}={\bf 0}$, is a $2$-dimensional 
invariant subspace where passive advection takes place.

\section{The open flow model}

In order to illustrate the 
finite-size effects in open chaotic advection,
we choose the blinking vortex-source system for the flow \cite{karolyi1,aref2}.
This system is periodic and consists of two 
alternately open point sources in a plane.
It models the alternate injection of rotating fluid
in a large shallow basin.
Apart from the two point sources, the 
dynamics is Hamiltonian, with
the streamfunction given by
$\Psi=-\left(K\ln r'+Q\phi'\right) \Theta(\tau)
-\left(K\ln r''+Q\phi''\right) \Theta(-\tau)$,
where $\tau=0.5T - t \mbox{ mod }T$. 
Here, $r'$ and $\phi'$ are
polar coordinates centered at $(-1,0)$ whereas
$r''$ and $\phi''$ are
polar coordinates centered at $(1,0)$.
The two parameters $Q$ and $K$ are, respectively, the strenghts of the
source and of the vortex. The period of the flow is $T$, whereas
$\Theta$ stands for the Heaviside step function.
The sources are located at positions $(\pm 1,0)$. 
For each half period, the system remains stationary with only 
one of the sources open. This allows one to 
analytically integrate the equations of motion, 
$u_x=\partial \Psi /\partial y$ and
$u_y=-\partial \Psi /\partial x$,
during each half period,
and 
thus to write explicitly a stroboscopic map 
for this system, recording the positions of the particles
after integer multiples of the period $T$. 
In complex representation, $z=x+iy=r \mbox{ exp}(i\phi)$, 
the stroboscopic map is 
\begin{equation}
z_{n+1}=\left(z'_n-1\right)
\left(1-\frac{\eta}{|z'_n-1|^2}\right)^{1/2-i\xi/2}+1,
\label{estroboscopico}
\end{equation}
where 
$z'_n=\left(z_n+1\right)
(1-\frac{\eta}{|z_n+1|^2})^{1/2-i\xi/2}-1,$
and where the two parameters governing the dynamics are 
$\eta\equiv QT$ and $\xi\equiv K/Q$. We fix 
$\eta=-0.5$ and $\xi=10$, for which the dynamics is 
chaotic. 
Figure 1(a) shows the unstable manifold of the
chaotic saddle for passive tracers.

\section{Discrete description of the dynamics}

Equation (\ref{estroboscopico}) is the discrete 
version of the flow whose streamfunction is $\Psi$.
Thus, passive advection is described by
the two-dimensional area preserving 
map ${\bf x}_{n+1}={\bf f}({\bf x}_n)$,
in our case the one given by Eq. (\ref{estroboscopico})
with ${\bf x}_n$ written in complex representation as $z_n$.
Analogously, a discrete version of Eq. (\ref{babiano}) for 
the dynamics of finite-size particles is given by \cite{cartwright1}
\begin{equation}
{\bf x}_{n+2}-{\bf f}({\bf x}_{n+1})=\mbox{e}^{-\gamma}\mathsf{J}_{\bf f}({\bf x}_n)\left({\bf x}_{n+1}-{\bf f}({\bf x}_n)\right).
\label{bailout}
\end{equation}
Equation  (\ref{bailout})
contains all the essential features of the flow 
given by Eq. (\ref{babiano}).
In particular,
both have constant rate of phase space contraction.
The term $-(\mathsf{J}_{\bf u}+\frac{2}{3\mbox{ St}}\mathbbm{1})$
in Eq. (\ref{babiano})
is substituted 
by the expected term 
$\mbox{e}^{-\gamma}\mathsf{J}_{\bf f}({\bf x}_n)$
in the map. The number
$e^{-\gamma}$ plays the role 
of the Stokes parameter in the discrete dynamics.
Equation (\ref{bailout}) has been successfully used in 
the study of the dynamics of neutrally buoyant finite-size particles
under chaotic advection \cite{cartwright2} and it was generalized
in order to describe also the dynamics of particles whose density differs
from that of the fluid \cite{motter1}. 
It is convenient to 
cast Eq.  (\ref{bailout}) into 
\begin{equation}
\left\{ \begin{array}{l}
{\bf x}_{n+1}={\bf f}({\bf x}_n)+{\bf w}_n,  \\[0.3cm] 
{\bf w}_{n+1}=\mbox{e}^{-\gamma}\mathsf{J}_{\bf f}({\bf x}_n){\bf w}_n.
\end{array} \right.   \label{bailout2}
\end{equation}
If $\lambda_{\bf x}$ and $\lambda_{\bf x}^{-1}$
are the eigenvalues of $\mathsf{J}_{\bf f}({\bf x})$,
the vector ${\bf w}_n={\bf x}_{n+1}-{\bf f}({\bf x}_n)$
is amplified in regions where 
$\gamma <|\ln |\lambda_{\bf x}||$.
This corresponds to regions of the flow
characterized by larger strain.
In open chaotic flows with asymptotic
regularity, such regions are precisely 
where the chaotic saddle is located.
Consequently, there the finite-size particles
may detach from their corresponding fluid
elements and it is where 
the finite-size effects
are expected to occur.

\begin{figure}
\includegraphics[angle=0,width=8cm,height=4.3cm]{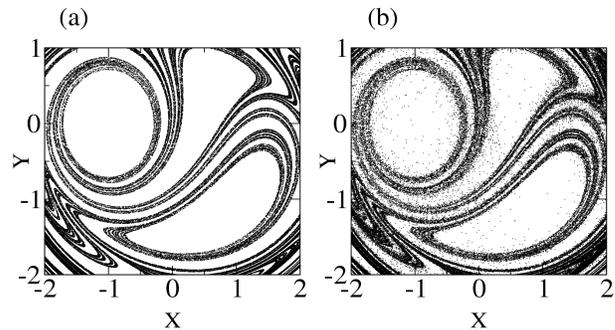}
\label{fig1} 
\caption{
(a) Unstable manifold of the chaotic saddle.
(b) The projection onto configuration space of the seventh
iteration of Eq. (\ref{bailout2}) for an ensemble of 
particles with $\mbox{e}^{-\gamma}=0.2$.
Initially, $10^6$ particles are uniformly distributed
in the hypercube 
$[-0.5,0.5]\times[-1,0]\times[-10^{-3},10^{-3}]\times[-10^{-3},10^{-3}]
\subset\mathbb{R}^4$.
We plot only the region $[-2,2]\times[-2,1]$, 
where about $9\times10^4$ are mapped to. 
}
\end{figure}

\section{Dispersion of the ensemble of finite-size particles}

Analogously to the continuous case, 
the dynamics described by 
Eq. (\ref{bailout2}) 
takes place in a $4$-dimensional phase space. 
The configuration space corresponds
to the $2$-dimensional invariant subspace 
${\bf w}={\bf 0}$, where passive advection occurs.
The limit of vanishing particle size
corresponds to $\mbox{St} \rightarrow 0$ in the flow and to 
$e^{-\gamma} \rightarrow 0$ in the map.
The solutions in this case are, respectively, ${\bf v}={\bf u}$
and ${\bf x}_{n+1}={\bf f}({\bf x}_n)$.
In this limit, we recover the motion of 
passive advection and an ensemble of particles 
initially located in the region of the chaotic saddle traces
asymptotically its unstable manifold in the configuration space.
As the size of the particles grows from zero, 
however, the particles trace out
the unstable manifold
in the $4$-dimensional phase space.
The key point is that, when projected onto the
configuration space, they do not trace 
the unstable manifold, $W^u(\Sigma)$, 
corresponding to passive advection,
but they instead ``disperse'' about it. 
Figure 1(b) shows 
the projection onto the $2$-dimensional configuration space 
of the seventh iteration of an ensemble of finite-size
particles, initially placed in the
region of the chaotic saddle in the $4$-dimensional phase space. 
We thus have a clear qualitative picture of the 
finite-size effects, in which the smearing
of the fractal structure in configuration space is evident. 
As the size of the
particles gets larger, their ``dispersion''  
about $W^u(\Sigma)$
increases.

\section{Theory for the dispersion}

To obtain a quantitative description of the phenomenon,  
we define 
the {\it dispersion}
$D$ of a set $S_1$ about a 
set $S_2$ as the average of the distances $d(x,S_2)$
between the points $x \in S_1$ and the set  
$S_2$.
So we have 
\begin{equation}
D = \langle d(x,S_2)\rangle _{x\in S_1}, \label{definition}
\end{equation}
where $d(x,S_2)=\mbox{min}\{d(x,y),y\in S_2\}$.
Our goal is the derivation of 
an expression predicting the behaviour of
the dispersion, about $W^u(\Sigma)$,
of the 
position vectors in configuration space of an ensemble
of finite-size particles whose initial conditions are
$({\bf x}_0,{\bf w}_0)$,
as a function 
of $\mbox{e}^{-\gamma}$.
Differently from the case illustrated in Fig. 1(b),
the initial positions ${\bf x}_0$ are chosen in the chaotic saddle,
$\Sigma$,
as our aim is to 
make a theory on the invariant set.
Figure 2 shows the chaotic saddle.
It is carefully obtained to warrant the correct natural measure \cite{jacobs1}.
The initial conditions ${\bf w}_0$ are chosen so that
$\|{\bf w}_0\|\ll 1$, corresponding to 
${\bf v}\simeq{\bf u}$ in the flow, 
a condition assumed in the derivation of Eq.  (\ref{maxey}).
The directions of the vectors ${\bf w}_0$ are randomly chosen
with uniform angular distribution.
We calculate the dispersion for the second
iterate of the map,
as this is the lowest iterate where the finite-size effects
are present.  
The first iterate of the map, ${\bf x}_1$, is $\|{\bf w}_0\|$-distant
from ${\bf f}({\bf x}_0)$, independently of ${\mbox e}^{-\gamma}$,
as we can see readily from Eq. (\ref{bailout2}).
Figure 3 illustrates the situation that we aim to 
quantify.

\begin{figure}
\includegraphics[angle=0,width=6.8cm,height=15.3cm]{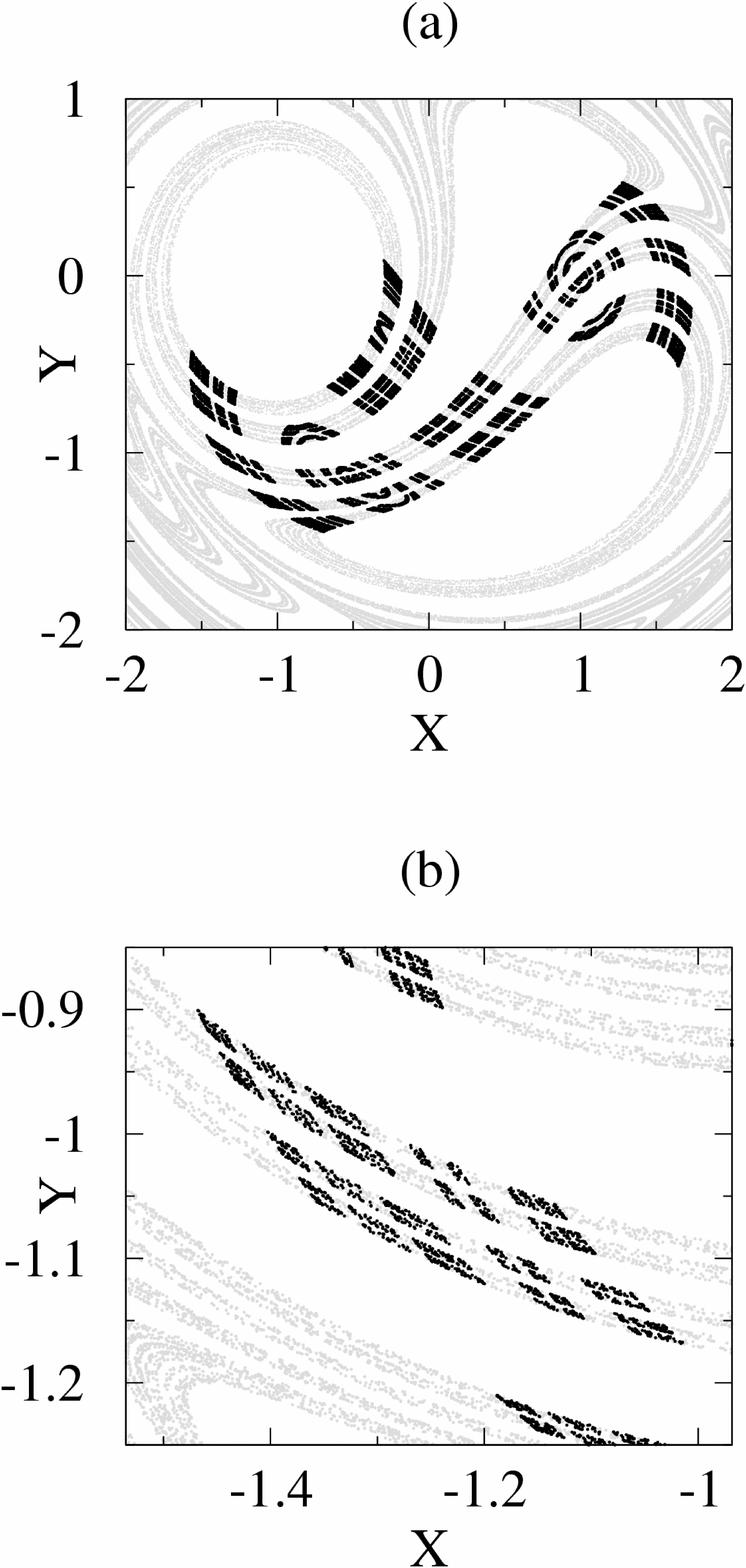}
\label{fig2} 
\caption{
(a) The chaotic saddle (black) and its unstable manifold (grey).
(b) Magnification of (a).
}
\end{figure}

\begin{figure}
\includegraphics[angle=0,width=6.8cm,height=15.3cm]{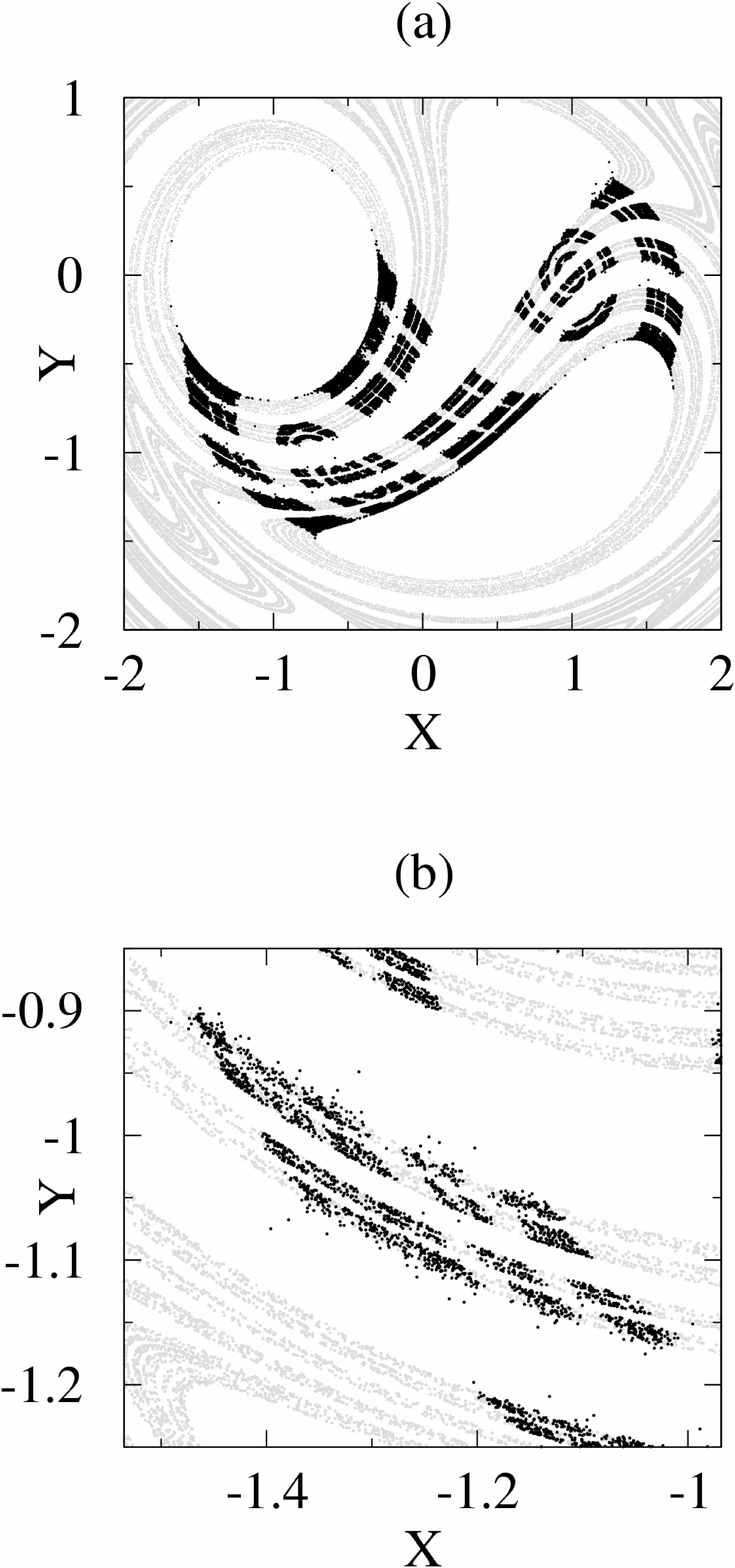}
\label{fig3} 
\caption{
(a) The projection onto configuration space of the second 
iteration of Eq. (\ref{bailout2}) for an ensemble of 
particles with $\mbox{e}^{-\gamma}=0.2$.
Initially, $6\times 10^4$ particles (black points) are 
placed in the chaotic saddle and the initial vectors
${\bf w}_0$ are uniformly
chosen in the range $0<\|{\bf w}_0\|<10^{-3}$, 
and with
uniform angular distribution.
The unstable manifold $W^u(\Sigma)$ is also shown (grey points).
(b) Magnification of (a).
}
\end{figure}

In order to obtain the dispersion $D$ of the ensemble of 
finite-size particles about $W^u(\Sigma)$,
we follow two steps. 
First, we derive an expression for the expected value, 
over both the natural measure of the chaotic saddle and 
the distribution of the initial vectors ${\bf w}_0$,
of the distance between ${\bf x}_2$
and the unstable subspace arising from 
${\bf f}^2({\bf x}_0)$.
This expected value $\delta_{\gamma}$
depends on the Stokes parameter $\mbox{e}^{-\gamma}$.
The derivation of $\delta_{\gamma}$ involves only
dynamical arguments.
Second, we derive an expression for $D$ as 
a function of $\delta_{\gamma}$.
This expression is necessary because the 
value of $D$ is smaller than $\delta_{\gamma}$.
$W^u(\Sigma)$ is a fractal set and
usually contains curves
whose expected distance to the vector positions of the finite-size 
particles is smaller than 
$\delta_{\gamma}$.
The situation is sketched in Fig. 4.
Accordingly, the correct value of the dispersion
$D$ of the finite-size
particles about $W^u(\Sigma)$ is given by the expression
\begin{equation}
D=F(\delta_{\gamma}), \label{formula}
\end{equation}
which defines the function $F$.
This function depends only on the geometry of the
set $W^u(\Sigma)$ in the region close to the
chaotic saddle. 
While an analytical expression for $F$
is a rather nontrivial task, 
this problem can be overcomed by 
means of a computer-aided approach.

Let us now derive the expression for $\delta_{\gamma}$.
Projected onto configuration space, the second iterate
of the map, to first order in ${\bf w}_0$, is given by
\begin{equation}
{\bf x}_2={\bf f}({\bf x}_1^{*})+\mathsf{J}_{\bf f}({\bf x}_1^{*}){\bf w}_0
+\mbox{e}^{-\gamma}\mathsf{J}_{\bf f}({\bf x}_0){\bf w}_0,
\label{second}
\end{equation}
where ${\bf x}_1^{*}={\bf f}({\bf x}_0)$.
The term 
$\mathsf{J}_{\bf f}({\bf x}_1^{*}){\bf w}_0$ 
corresponds to a point of the ellipse centered at
the origin and having axes
of lenghts equal to $\sqrt{\lambda_{1i}}\|{\bf w}_0\|$, where
$\lambda_{1i}$, $i=1,2$, are the eigenvalues of
$\mathsf{J}_{\bf f}({\bf x}_1^{*})\mathsf{J}_{\bf f}^{T}({\bf x}_1^{*})$.
In an analogous fashion, the term
$\mbox{e}^{-\gamma}\mathsf{J}_{\bf f}({\bf x}_0){\bf w}_0$
corresponds to a point of the ellipse centered at
the origin and having axes 
whose lenghts are  
$\mbox{e}^{-\gamma}\sqrt{\lambda_{0i}}\|{\bf w}_0\|$,
where $\lambda_{0i}$, $i=1,2$, are the eigenvalues of
$\mathsf{J}_{\bf f}({\bf x}_0)\mathsf{J}_{\bf f}^{T}({\bf x}_0)$.
Thus we see that the right-hand side of Eq. (\ref{second})
corresponds to a sum of two position vectors located at ellipses
centered at ${\bf f}({\bf x}_1^{*})$.

Now, because of the large strain in the region of 
the chaotic saddle,
the average over its natural measure
of the lenght of the
major axis of the ellipse 
$\mathsf{J}_{\bf f}({\bf x})N$, where $N$
is the unit disk in $\mathbb{R}^2$,
is expected to be much larger than unit.
This fact implies that the major axis
of the ellipse 
$\mathsf{J}_{\bf f}({\bf x}_1^{*})N$
is, in general,  approximately colinear with 
the unstable subspace $E^u ({\bf f}({\bf x}_1^{*}))=E^u ({\bf f}^2({\bf x}_0))$
arising from ${\bf f}^2({\bf x}_0)$.
To understand that, 
consider the unit disk $N$ centered at ${\bf x}_1^{*}$.
Let $\hat{a}_2$ and $\hat{b}_2$ be the unit vectors in
the directions, respectivelly, of the  
major and minor axes of the ellipse 
$\mathsf{J}_{\bf f}({\bf x}_1^{*})N$.
Let $\hat{a}_1$ and $\hat{b}_1$ be the unit vectors 
in the directions of the pre-images of $\hat{a}_2$ and 
$\hat{b}_2$, respectively. 
Now let $\hat{u}_1$ be the unit vector in the direction
of the unstable space of ${\bf x}_1^{*}$.
We can write $\hat{u}_1=c_{1}\hat{a}_1+c_{2}\hat{b}_1$,
with $c_{1}$ and $c_{2}$ uniquely determined, 
as $\hat{a}_1$ and $\hat{b}_1$ form a basis in the plane.
We then have 
$\mathsf{J}_{\bf f}({\bf x}_1^{*})\hat{u}_1=
c_{1}\sqrt{\lambda}\hat{a}_2+(c_{2}/\sqrt{\lambda})\hat{b}_2$,
where $\lambda=\mbox{max}\{\lambda_{11},\lambda_{12}\}$
(notice that $\lambda_{11}=1/\lambda_{12}$, as
${\bf f}$ is an area preserving map). 
The vector $\mathsf{J}_{\bf f}({\bf x}_1^{*})\hat{u}_1$ is, of 
course, in the same direction of the unstable subspace arising from the point
${\bf f}^2({\bf x}_0)$. 
Taking the scalar 
product $\mathsf{J}_{\bf f}({\bf x}_1^{*})\hat{u}_1\cdot\hat{a}_2$,
we obtain the angle $\theta$ between $\hat{a}_2$ and
the unstable subspace arising from ${\bf f}^2({\bf x}_0)$ as
$\theta=\arccos{[\left(1+c_{2}^2/(c_{1}^2\lambda^2)\right)^{-1/2}]}$.
Because $c_{2}$ is not expected to be much larger than $c_{1}$
and because of the large strain in the region of the chaotic
saddle, we typically have $|c_{2}/(c_{1}\lambda)|\ll 1$.
Thus, to second order in $1/\lambda$, we have $\theta=|c_{2}/(c_{1}\lambda)|$, 
and we see that 
the major axis
of the ellipse 
$\mathsf{J}_{\bf f}({\bf x}_1^{*})N$
is usually almost colinear with 
the unstable subspace arising from ${\bf f}^2({\bf x}_0)$.

The directions of the axes of the ellipse 
$\mbox{e}^{-\gamma}\mathsf{J}_{\bf f}({\bf x}_0)N$, 
on the other hand, do not depend on 
the direction of the unstable space 
at ${\bf f}^2({\bf x}_0)$. 
Figure 5 illustrates the situation.
Assuming that the 
distribution of the angles between the major axes of
the ellipses 
$\mathsf{J}_{\bf f}({\bf x}_1^{*})N$
and $\mbox{e}^{-\gamma}\mathsf{J}_{\bf f}({\bf x}_0)N$
is uniform 
over the natural measure of the chaotic saddle
(our theory  
remains a good approximation as long as this distribution is 
not too concentrated on a small range of angles) and 
choosing our reference frame centered at ${\bf f}^2({\bf x}_0)$ and
such that 
the x-axis is in the direction of the unstable subspace
arising from ${\bf f}^2({\bf x}_0)$,
the expected value for the 
distance between ${\bf x}_2$
and the unstable subspace arising from
${\bf f}^2({\bf x}_0)$ is given by
\begin{equation}
\delta_{\gamma}=\frac{1}{8\pi^3}\int_0^{2\pi}\int_0^{2\pi}\int_0^{2\pi}\abs{y}\mbox{d}\alpha 
\mbox{d}\beta \mbox{d}\omega,
\label{function}
\end{equation} 
where ${\bf r}=(x,y)=(\mathsf{A} + 
\mbox{e}^{-\gamma}\mathsf{R}_{\alpha}\cdot\mathsf{B}\cdot\mathsf{R}_{\beta}).{\bf r}_0$, 
$\mathsf{A}=\pmatrix {1/\langle\lambda^{-1/2}\rangle & 0\cr 0 & \langle\lambda^{-1/2}\rangle}$,
$\mathsf{B}=\pmatrix {\langle\lambda^{1/2}\rangle & 0\cr 0 & 1/\langle\lambda^{1/2}\rangle}$, 
$\mathsf{R}_{\alpha}$ and
$\mathsf{R}_{\beta}$ are the rotation matrices for the angles $\alpha$ and $\beta$, 
and ${\bf r}_0=\langle\|{\bf w}_0\|\rangle(\cos{\omega},\sin{\omega})$.
The averages $\langle \lambda^{-1/2}\rangle$ and 
$\langle \lambda^{1/2}\rangle$
are taken over the natural measure of the chaotic saddle.
The right-hand side of Eq. (\ref{function}) represents an average 
over  the directions of the position vectors $\mathsf{J}_{\bf f}({\bf x}_1^{*}){\bf w}_0$
(integral over $\omega$) 
and $\mbox{e}^{-\gamma}\mathsf{J}_{\bf f}({\bf x}_0){\bf w}_0$
(integral over $\beta$),
and over the direction of the major axis of the ellipse 
$\mathsf{J}_{\bf f}({\bf x}_0)N$
(integral over $\alpha$) (see Fig. 5).
The matrix $\mathsf{A}$ accounts for 
the effects of the term 
$\mathsf{J}_{\bf f}({\bf x}_1^{*}){\bf w}_0$.
Its contribution to the distance from the point ${\bf x}_2$ to the 
unstable subspace arising from
${\bf f}^2({\bf x}_0)$ 
depends on the minor axis of 
the ellipse $\mathsf{J}_{\bf f}({\bf x}_1^{*})N$.
For this reason the non-zero elements of $\mathsf{A}$
involve $\langle\lambda^{-1/2}\rangle$.
The matrix 
$\mbox{e}^{-\gamma}\mathsf{R}_{\alpha}\cdot\mathsf{B}\cdot\mathsf{R}_{\beta}$
describes the effects of the term
$\mbox{e}^{-\gamma}\mathsf{J}_{\bf f}({\bf x}_0){\bf w}_0$.
The contribution of this term to $\delta_{\gamma}$
depends essentially on the major axis of the 
ellipse $\mathsf{J}_{\bf f}({\bf x}_0)N$. This is why 
the non-zero elements of $\mathsf{B}$
involve $\langle\lambda^{1/2}\rangle$ (notice that 
$\langle\lambda^{1/2}\rangle \neq 1/\langle\lambda^{-1/2}\rangle$).

\begin{figure}
\includegraphics[angle=0,width=7cm,height=4cm]{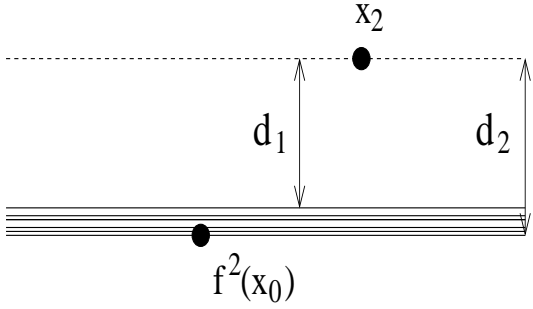}
\label{fig4} 
\caption{
Sketch of the position ${\bf x_2}$ in configuration space, and
its distances $d_1$ to  $W^u(\Sigma)$ and $d_2$ to the unstable subspace arising from
${\bf f}^2({\bf x}_0)$.
The dispersion $D$ is
the average of $d_1$, whereas $\delta_{\gamma}$ is the average of $d_2$. 
}
\end{figure}

\begin{figure}
\includegraphics[angle=0,width=7cm,height=4cm]{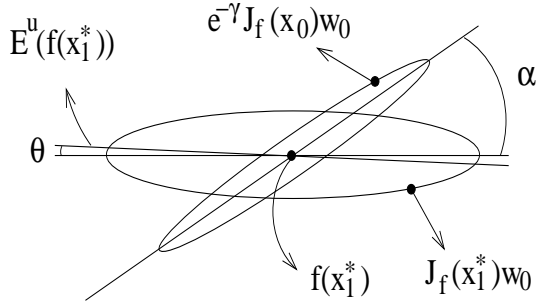}
\label{fig5} 
\caption{
Sketch of the situation leading to Eq. (\ref{function}).
}
\end{figure}

\section{Results}
 
We have computed $D$ using our theory, Eq. 
(\ref{formula}) (along with Eq. (\ref{function})).
The initial conditions ${\bf w}_0$ were uniformly
chosen in the range $0<\|{\bf w}_0\|<10^{-3}$, 
so that $\langle\|{\bf w}_0\|\rangle=5\times 10^{-4}$, and with
uniform angular distribution.
The result is shown in Fig. 6. The inset of Fig. 6
regards the estimation of the function $F$.
A good approximation for $F$ is
obtained by assuming that $F(\delta)$ is the dispersion, 
about $W^u(\Sigma)$, 
of a set of points which are a distance $\delta\pi/2$ apart
(in homogeneously distributed directions)
from the points in the chaotic saddle.
The factor $\pi/2$ 
is necessary
because, if $\delta\pi/2$
is the isotropically distributed distance to the point at 
the chaotic saddle, then
$\delta$
is the expected value for the distance to the unstable subspace
arising from that point. 

In order to validate our theory, we also measured numericaly
the dispersion,
directly using the definition, Eq. (\ref{definition}). 
The results are also presented in Fig. 6, for comparison, 
and they 
are very close to the ones predicted by our theory.

\begin{figure}
\includegraphics[angle=0,width=7cm,height=7cm]{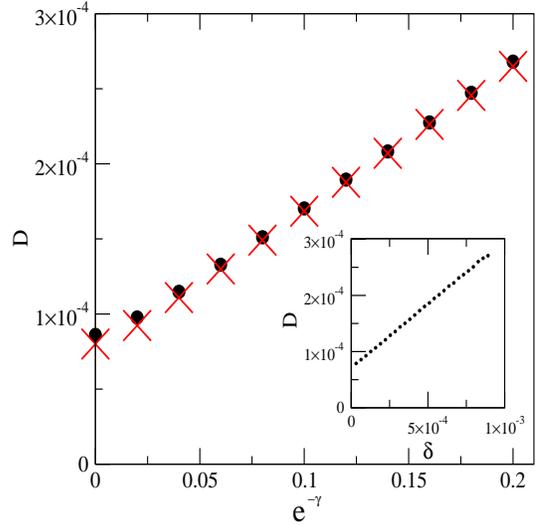}
\label{fig6} 
\caption{
(color online).
Theoretical (red $\times$ symbols) and
numerical (black circles)  
values of the dispersion $D$ 
of an ensemble of $6\times10^4$ 
finite-size particles about
$1.8\times 10^7$ points representing $W^u(\Sigma)$.
For the theoretical values, we use Eq. 
(\ref{formula}) (along with Eq. (\ref{function})).
The values $\langle \lambda^{-1/2}\rangle=0.12$ and  
$\langle \lambda^{1/2}\rangle =21.06$
were computed numerically.
The inset shows the dispersion $D$ of  
$6\times10^4$ 
randomly generated points (at a distance $\delta\pi/2$
from  
the points of the chaotic saddle) about the same
$1.8\times 10^7$ points representing $W^u(\Sigma)$.
Interestingly, the overall scaling is linear in the range 
$3\times 10^{-5} < \delta < 9\times 10^{-4}$.
By regression,
we obtain 
$F=7.128\times10^{-5}+2.266\times10^{-1}\times\delta$.
The fact that $F(0)\neq 0$ is, of course, due to our approximation of
$W^u(\Sigma)$ by a finite number of points.
}
\end{figure}

\section{Conclusions}

In summary,
we have derived a theory accounting for the quantitative
behaviour of the dispersion 
of neutrally buoyant finite-size particles about the unstable 
manifold of the chaotic saddle 
as a
function of the Stokes parameter.
Our theory refers to the onset of the 
finite-sizeness induced dispersion in 
the discrete description of the dynamics 
(second iterate of the map given by Eq. (\ref{bailout2})). 
The theory involves a dynamical
part, where the computation of the averages 
$\langle \lambda^{-1/2} \rangle$ and 
$\langle \lambda^{1/2} \rangle$ is required, 
and a geometrical part,
regarding the function $F$.  
This geometrical part will 
be the same for 
any theory of the dispersion, 
either for higher iterates of the map, Eq. (\ref{bailout2}),
or for snapshots in the continuous description, Eq. (\ref{babiano}).
It is a highly difficult problem to be solved analitically,
but it can be overcomed by means of a computer-aided approach.
The dynamical part of the theory will be different
for higher iterates of the map
or for snapshots in the continuous description, but it will be
based
on the same type of reasoning.

An interesting open problem is to determine
the restrictions that the dispersion 
predicted by our theory imposes
on the enhancement of activity (e.g. biological) \cite{toroczkai1}
of finite-size particles.

\begin{acknowledgments}

The authors thank
T. T\'el, 
A.E. Motter, 
S. Kraut, and K.M. Zan for
fruitful hints and discussions.
R.D.V. is specially grateful to 
P.A.S. Salom\~ao for 
illuminating discussions
and to L. Moriconi for a valuable suggestion.
This work was supported by FAPESP and CNPq.

\end{acknowledgments}

\end{document}